\documentclass[12pt]{iopart}
\usepackage{graphicx}
\usepackage{pxfonts}
\usepackage{color}
\begin{document}
\title[O and O$_3$ production in the effluent of a He/O$_2$ microplasma jet]{The mystery of O and O$_3$ production in the effluent of a He/O$_2$ atmospheric pressure microplasma jet}

\author{D Ellerweg, A von Keudell, and  J Benedikt}
\address{Institute for Experimental Physics II: Reactive Plasmas,\\Ruhr-Universit\"at Bochum, 44780 Bochum, Germany}

\ead{jan.benedikt@ruhr-uni-bochum.de}

\begin{abstract}
Microplasma jets are commonly used to treat samples in ambient air atmosphere. The effect of admixing air into the effluent may severely affect the composition of the emerging species. Here, the effluent of a He/O$_2$ microplasma jet has been analyzed in a helium and in an air atmosphere by molecular beam mass spectrometry.  First, the composition of the effluent in air has been recorded as a function of the distance to determine how fast air admixes into the effluent. Then, the spatial distribution of atomic oxygen and ozone in the effluent has been recorded in ambient air and compared to measurements in a helium atmosphere. Additionally, a fluid model of the gas flow with reaction kinetics of reactive oxygen species in the effluent has been constructed. In ambient air, the O density declines only slightly faster with the distance compared to a helium atmosphere. On the contrary, the O$_3$ density in ambient air increases significantly faster with the distance compared to a helium atmosphere. This mysterious behavior can have big implication for the use of similar jets in plasma medicine. It is shown that photodissociation of O2 and O3 is not responsible for the observed effect. A reaction scheme involving the reaction of plasma produced highly vibrationally excited O$_2$ with ground state O$_2$ molecules is proposed as a possible explanation of the observed densities. A very good agreement between measured and simulated densities is achieved.
\end{abstract}

\submitto{\PSST}

\maketitle

\section{Introduction}

Cold atmospheric plasmas exhibit many unique properties, which make them very attractive for a
broad field of applications. These plasmas offer non-equilibrium chemistry at atmospheric pressure
with high densities of reactive neutral and charged species and high fluxes of energetic photons.
They can be utilized as light sources, for localized treatment of thermo and vacuum sensitive
materials, and, especially in the case of jets, also of living tissue in plasma medicine
applications.
\\Cold atmospheric plasma jets have been developed and studied in the past for many different
applications, like thin film deposition \cite{Schafer2008,Benedikt2007,Raballand2009}, inactivation
of pathogens \cite{Goree2006,Rahul05,Perni2007,Sladek2005a} or generation of nanoparticles
\cite{Chiang2007}. Many of the developed sources use helium as plasma forming gas \cite{Walsh2010,
Knake08, Karakas2010, Reuter2011} with a small admixture of a molecular gas (e.g. molecular oxygen). It
has been shown that these jets are effective in producing reactive oxygen species (ROS) and can
effectively inactivate for example bacteria or etch hydrocarbon polymer layers \cite{Schneider2011}.
\\For the understanding of the interaction of the plasma effluent with substrates, of inactivation of
bacteria, or of its effects on living tissues, it is necessary to know absolute densities and fluxes
of reactive species in the plasma jet. Moreover, the microplasma jets are in many cases operated in
ambient atmosphere, which effect on the plasma effluent is not fully and quantitatively understood.
\\Measurements of absolute densities of reactive species in microplasma jets are a nontrivial task due
to the high pressure and the small dimensions. Advanced diagnostics like two-photon absorption
laser-induced fluorescence spectroscopy (TALIF) \cite{Knake08,Knake2008,Ellerweg2010}, molecular
beam mass spectrometry (MBMS) \cite{Ellerweg2010} or absorption spectroscopy \cite{Sousa2008} have
to be used to obtain absolute ROS densities. It is important for possible applications that all
measurements are carried in realistic situations. Many experiments are, however, performed in a
controlled atmosphere of noble gas, whereas possible applications will be performed in ambient air.
\\Here, absolute atomic oxygen and ozone densities in the effluent of a He/O$_2$ micro-scaled
atmospheric pressure plasma jet ($\mu$-APPJ) operated in ambient helium and air are measured by
absolutely calibrated MBMS. The main focus of the work presented here is to analyze the influence
of ambient air on the ROS densities and to reveal possible reaction pathways leading to their
production. This is achieved by a comparison of the experimental data to a fluid model of the gas
flow combined with chemical reactions in the plasma effluent.

\section{Experimental setup}

Only a brief description of the experimental setup is provided here since it has been
described in detail elsewhere \cite{Ellerweg2010,Benedikt2009}. The main difference is that in this
study the $\mu$-APPJ has been operated and analyzed in ambient air in contrast to previous measurements in an
ambient helium atmosphere.

\subsection{Micro-scaled atmospheric pressure plasma jet}
\begin{figure}
    \centering
        \includegraphics[width=0.7\textwidth]{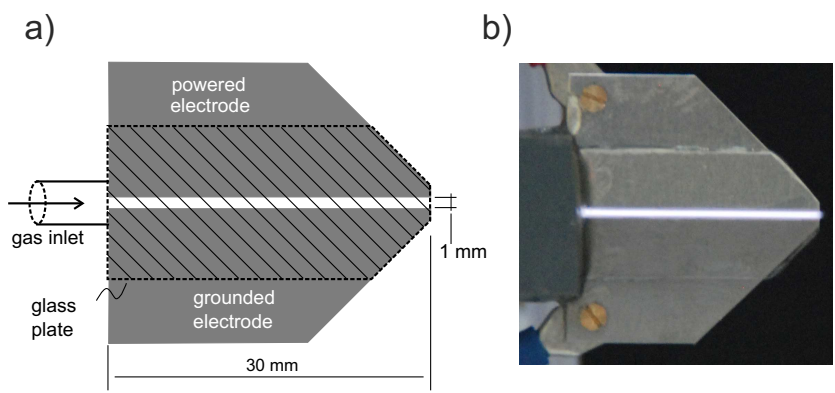}
    \caption{Sketch (a) and photo (b) of the $\mu$-APPJ}
    \label{fig:APPJ}
\end{figure}

A schematic sketch of the already defined $\mu$-APPJ is shown in
figure \ref{fig:APPJ}. The microplasma jet consists of two parallel stainless steel electrodes
(30\,mm long, 1\,mm thick) separated by a gap of 1\,mm. One electrode is powered by an rf power
supply (13.56\,MHz, absorbed power $<$1\,W), while the other one is grounded. The $\mu$-APPJ is
operated at a gas flow of 1.4\,slm helium with a small admixture of molecular oxygen ($<$1.6\,\%).
The plasma volume of 30x1x1\,mm$^3$ is confined on both sides by glass plates.
\\If not stated otherwise, the plasma is operated with a gas flow of 1.4\,slm He with 0.6\,\% O$_2$
and an applied electrode voltage of 230\,V$_{RMS}$.
\\Previous measurements in a He atmosphere \cite{Knake08, Knake2008, Ellerweg2010} revealed that the
$\mu$-APPJ is very effective in producing ROS. An O$_2$ dissociation rate up to 20\% and maximum O
and O$_3$ densities around 10$^{15}$cm$^{-3}$ could be measured. The O density is maximal at an
O$_2$ admixture of 0.6\,\%, while the O$_3$ density continuously increases with the O$_2$
admixture. The atomic oxygen density increases when a higher power is applied to the $\mu$-APPJ,
whereas the ozone density slightly decreases. With increasing distance from the $\mu$-APPJ nozzle,
the O concentration decreases and the O$_3$ concentration increases.

\subsection{X-Jet}
\begin{figure}
    \centering
        \includegraphics[width=0.7\textwidth]{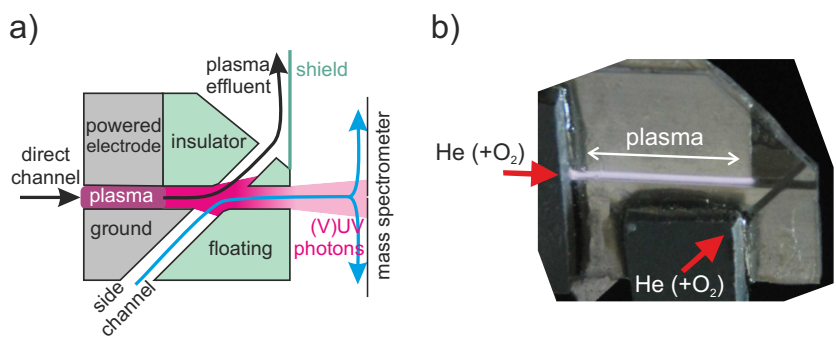}
    \caption{Sketch (a) and photo (b) of the X-Jet}
    \label{fig:x-jet}
\end{figure}

The so-called X-Jet is a modified version of the $\mu$-APPJ and allows the separation of heavy
species from photons in the effluent downstream of the plasma. The separation is realized in the
crossed channel structure at the nozzle of the jet, where a direct channel (extension of the space
between electrodes) is crossed by an additional side channel (cf. figure \ref{fig:x-jet}). A helium
flow introduced through the side channel of this structure deflects the particles in the plasma
effluent (incl. radicals and metastables) from their movement through the direct channel into a
side channel. The photons, on the contrary, are not affected by this additional gas flow and they
can propagate undisturbed through the direct channel. A detailed description of the X-Jet and tests
of its performance can be found in the literature \cite{Schneider2011, Schneider2011b}. As will be
demonstrated later, the X-Jet allows to test the photochemistry induced by plasma generated photons in
the gas mixture introduced through the side channel (cf. figure \ref{fig:x-jet}\,a) and in ambient air.

\subsection{Molecular beam mass spectrometer}
\begin{figure}
    \centering
        \includegraphics[width=0.4\textwidth]{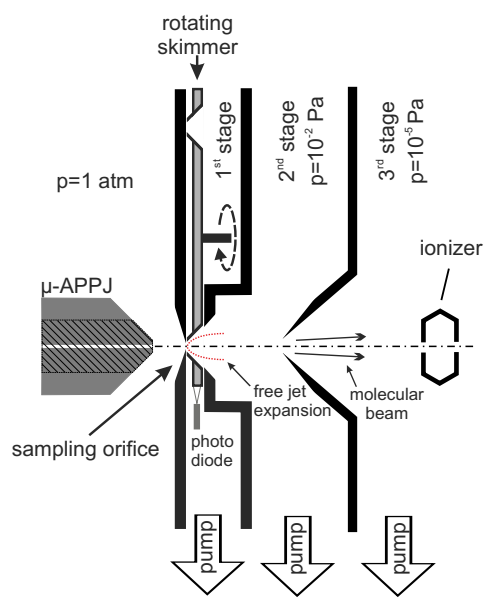}
    \caption{Schematic of the differential pumping system}
    \label{fig:MBMS}
\end{figure}

\begin{figure}
    \centering
        \includegraphics[width=0.5\textwidth]{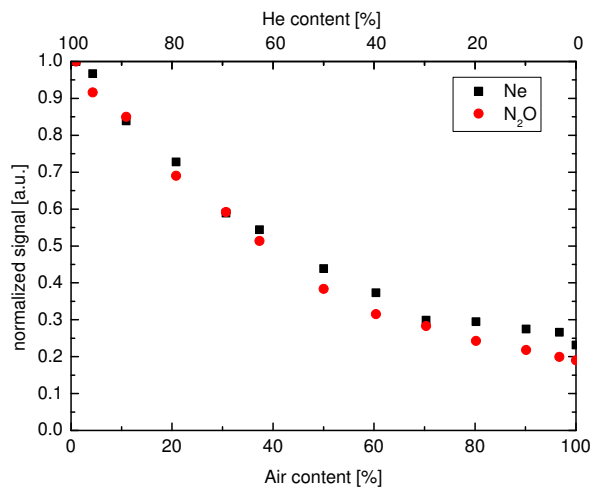}
    \caption{Measurement of the Ne and N$_2$O signal of different He/air mixtures with constant addition of Ne (1\%) and N$_2$O (0.5\%)}
    \label{fig:NeN2O}
\end{figure}

A molecular beam mass spectrometer (MBMS) is used to study the effluent of the $\mu$-APPJ. Because
the operation of a mass spectrometer (MS) requires a good vacuum, a differential pumping system has
to be used. The system has been described in detail previously \cite{Benedikt2009}. In summary, the
MBMS consists of three pumping stages that are connected by small orifices (cf. figure
\ref{fig:MBMS}). The gas extraction from the atmosphere into the first pumping stage is done by a
sampling orifice with a diameter of 100\,$\mu$m. In the first stage, a special beam chopper (i.e.
rotating flat metal disk with four small embedded skimmers) is mounted to allow only a pulsed gas
flow into the differential pumping system. The second and third pumping stages are interconnected
via a skimmer ($\varnothing$ 0.8\,mm). The ionizer of the MS is installed in the third stage in
line-of-sight with the orifice and the skimmer to ensure that the molecular beam (MB) forming
behind the sampling orifice can reach the ionizer without any disturbances.
\\Two different MBMS designs have been used. On the one hand, the previously described one
\cite{Benedikt2009} with the front plate of the MBMS oriented vertically like shown in figure
\ref{fig:MBMS} and on the other hand a new MBMS setup with slightly different dimension and a
horizontally oriented front plate. The axis of the $\mu$-APPJ has always been placed normal to the
sampling orifice pointing directly at it (see figure \ref{fig:MBMS}).
\\A chamber for a controlled atmosphere is mounted at the front plate of the MBMS system and encloses
the plasma jet. The volume of the chamber (approx. 1.1\,dm$^3$) can be filled with helium or air.
\\The MS measures the species density in the ionizer. With a proper background correction, the
composition of the MB in the ionizer can be determined. When gas is sampled from atmosphere to low
pressure, the composition of the MB is, however, different from the original composition of the gas
mixture at the sampling orifice. This is due to the so-called composition distortion in MBMS
sampling. A supersonic expansion takes place downstream from the sampling orifice and a supersonic
free jet is formed \cite{Scoles1988a}. Several different effects are responsible for the change in
composition \cite{Knuth95}: radical recombination at external probe surface, acceleration into
probe orifice, chemical relaxation in free jet, radial diffusion in free jet, skimmer interference
and mach-number focusing. Probably, one of the most important effects is radial diffusion in the
supersonic free jet. Huge radial pressure gradients exist just behind the sampling orifice which
results into larger diffusion fluxes in the radial direction. The extent of this diffusion is
different for different species, depending mainly on their mass, collision cross section and main
collisional partner. Hence, pressure diffusion leads to mass separation downstream from the
orifice. It has been shown that heavy species tend to remain on the central streamline because
their diffusion is slow, whereas light species diffuse faster outwards along the pressure
gradients. This effect leads to discrimination of light species against heavy ones in the
composition of the MB.
\\The influence of composition distortion is difficult to determine theoretically because it consists
of many different effects. However, we can demonstrate it experimentally and use the observed
changes in the signal to compensate for the composition distortion.
\\The composition distortion is demonstrated by measuring the MS signals of a small amount of Ne (1\,\%)
or N$_2$O (0.5\,\%) in a He/air gas mixture. The air concentration in He/air mixture has been
varied between pure He (light atoms with small collision cross section as main collision partners)
and pure air (heavy diatomic molecules with larger collision cross section), whereas the Ne and
N$_2$O concentrations were kept constant. Ne is in pure He the heavier species and will enrich on
the center line of the MB. While changing the He/air ratio towards pure air, Ne becomes the lighter
species compared to the mean molecular mass of the mixture and will consequently diffuse faster out
of the MB. Figure \ref{fig:NeN2O} shows the relative Ne signal measured by the MBMS as function of
the air content in the He/air mixture. The Ne signal decreases by a factor of 4 during the He/air
ratio variation. The N$_2$O signal exhibits surprisingly a very similar behavior, even if it is
more than twice heavier than Ne and heavier as N$_2$ and O$_2$ molecules in air. The difference
between the masses of Ne, N$_2$O and air (N$_2$, O$_2$) is probably not significant and the effect
is dominated by their difference to He.
\\The measurements in the figure \ref{fig:NeN2O} show clearly that the composition distortion has
to be taken into account. The amount of air in the effluent and the effect of composition
distortion will strongly depend on the distance from the jet nozzle. The following procedure is
adopted to take this effect into account in the signal calibration.  First, the amount of air in
the effluent at a given distance from the jet is determined by measuring the He and N$_2$/O$_2$
signals. These signals are compared to the measurements obtained with He/air mixtures with well
defined composition. The local air content in the effluent can be determined in this way. This air
content, together with the data in the figure \ref{fig:NeN2O}, is used to determine the correction
factors necessary to compensate for the difference in the composition distortion in pure He and in
He/air mixtures. The O and O$_3$ signals are hence scaled up to the values, which would be obtained
in pure He. The additional calibration is performed as described previously for the measurements in
pure He \cite{Ellerweg2010}. Just briefly, O and O$_3$ is calibrated by comparing their signals
measured at masses 16 and 44 to signals measured for known densities of Ne and N$_2$O in He,
respectively. These gases have been chosen because of their similar mass and structure in
comparison to O and O$_3$. The electron impact ionization cross sections and the known mass
dependence of the transmission function of the MS for different species are taken into account as well.
Atomic oxygen has been measured with an electron energy in the ionizer of about 15\,eV to prevent
dissociative ionization of molecular oxygen. Ozone has been measured with an electron energy of
70\,eV because the signal does not overlap with dissociative ionization of other species.

\section{Fluid model of plasma effluent}
Absolute atomic oxygen and ozone densities in the effluent of the $\mu$-APPJ obtained by MBMS can
be used to reveal possible reaction pathways leading to their formation. The densities of ROS in
the effluent of similar plasmas have already been measured and the chemical kinetics of these
species have been modeled in the past \cite{Jeong2000,Reuter2008}.
\\The reaction pathways proposed in these works are used here as a starting reaction scheme used in a
2D axially symmetric fluid model of our $\mu$-APPJ. The model is based on the model already
described in our previous work \cite{Schneider2011}, however, the effect of ambient air is added
now and the measured absolute densities of O and O$_3$ are used to validate the model results. The
2D axially symmetric geometry is used to be able to model the diffusion of air into the gas stream
emanating from the plasma jet and to take the diffusion of ROS into account. The fluid model
combines the solution of the Navier-Stokes equations for the gas flow in the gas mixture of He and
air with the model of chemical kinetics of ROS. The model is solved using commercial COMSOL 3.5
software.

\subsection{Model of the gas flow}
\begin{figure}
    \centering
        \includegraphics[width=0.8\textwidth]{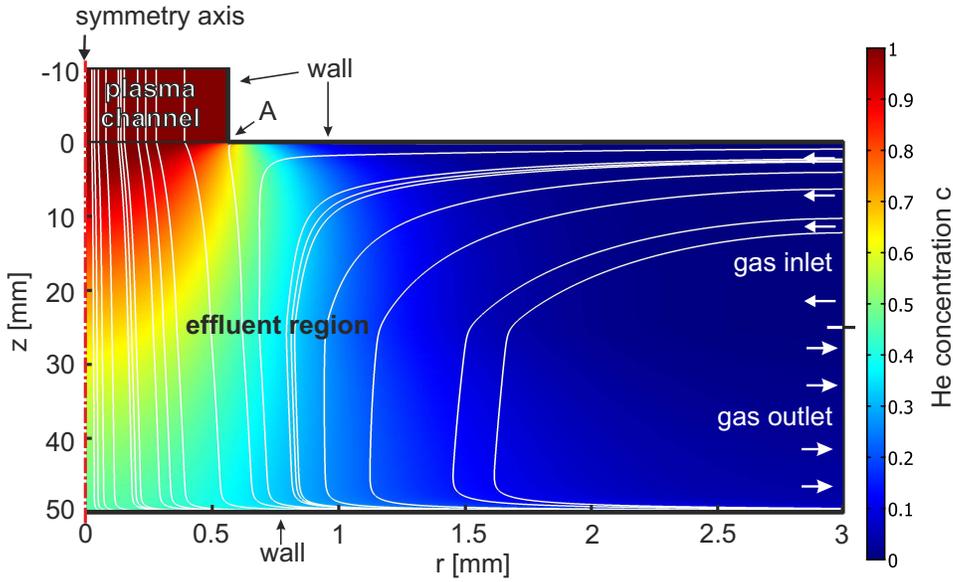}
    \caption{Geometry used in the model with He concentration $c$ in air and gas flow streamlines (white).
Note the different scaling of r- and z-axis.}
    \label{fig:geometry}
\end{figure}

The gas flow through the jet and in the effluent is solved first. Only He and air (represented by
N$_2$, see later) are considered in the model of the gas flow, because the concentration of O$_2$,
which is injected into the plasma, and all other plasma products is less than 1\,\% and can be neglected. The gas flow
is described by incompressible momentum conservation and continuity equations:

\begin{eqnarray}
\quad \rho \frac{\partial\textbf{u}}{\partial t} - \nabla \cdot\eta(\nabla\textbf{u} + (\nabla\textbf{u})^T) + \rho \textbf{u} + \nabla p = 0 \\
\quad \nabla \cdot \textbf{u} = 0 \label{Model_F}
\end{eqnarray}
with $\rho$ being gas density, $u$ gas flow velocity, $\eta$ dynamic viscosity and $p$ the pressure
(101325 Pa). No volume force is assumed to work on the gas.
\\The transport of He through the ambient air is simulated as a diffusion-convection transport:
\begin{equation}
\frac{\partial c}{\partial t} + \nabla \cdot(-D_{He}\nabla c) = - \textbf{u} \nabla c \label{Model_Dc}
\end{equation}
with c being the He concentration in the gas mixture ($c \in [0,1]$) and $D_{He}$ being the diffusion coefficient of He
in the gas mixture. The gas density, dynamic viscosity and also the diffusion coefficient are a
function of the He concentration, which varies in space due to air diffusion from the sides into
the effluent. The viscosity, density and diffusion coefficients of N$_2$ instead of air are used in
the flow simulation for simplicity. The density of the He-N$_2$ gas mixture is calculated as:
\begin{equation}
\rho_{mixture} = \rho_{He} \cdot c + \rho_{N_2} \cdot (1-c)
\end{equation}
where $c$ is the He concentration, $\rho_{He}$ = 0.164\,kg/m$^3$ and $\rho_{N_2}$ =
1.146\,kg/m$^3$. The viscosity of the He-N$_2$ mixture measured at 303.15\,K is taken from the literature
\cite{Brokaw1968} and fitted with following function:
\begin{equation}
\eta(c) =
-7.912\times10^{-6}\cdot c^4+1.154\times10^{-5}\cdot c^3-4.906\times10^{-6}\cdot c^2+3.355\times10^{-6}\cdot c+1.8\times10^{-5}
\end{equation}
The diffusion coefficients of He in the He-N$_2$ gas mixture is calculated using following formula:
\begin{equation}
\frac{1}{D_{He\,in\,mixture}} =\frac{c}{D_{He\,in\,He}}+\frac{1-c}{D_{He\,in\,N_2}}
\label{Model_Diff}
\end{equation}
A self diffusion coefficient of He in He $D_{He\,in\,He}$ = 1.7513$\times10^{-4}$\,m$^2$s$^{-1}$
and a diffusion coefficient of He in N$_2$ $D_{He\,in\,N_2}$ = 0.7337$\times10^{-4}$\,m$^2$s$^{-1}$
is used \cite{Pavlin, Poling}.
\\The jet and its effluent is modeled as two cylinders representing plasma channel and the effluent
region as indicated in figure \ref{fig:geometry}. The upper cylinder corresponds to the last 10\,mm
of the plasma channel and has a radius r=0.564\,mm (1\,mm$^2$ cross-section area). The top base
serves as a gas inlet, where a parabolic velocity profile through cylindrical tube calculated from
the gas flow of 1.4\,slm through 1\,mm$^2$ area (average velocity $\sim$25.6 m/s) and He
concentration equal one are used as boundary conditions. The bottom base serves as a gas outlet
from the plasma channel into the effluent. No slip for the gas flow and insulation/symmetry for
the He transport are selected as a boundary conditions for the outer wall of this cylinder.
\\The effluent is modeled as a second cylinder with 3\,mm radius and length of 50\,mm. The bottom
base and the outer part of the top base of this cylinder are assumed to be solid with no slip
boundary condition. The outer wall is divided into two parts. The lower half of its length is
defined as a gas outlet (boundary condition: convective flux, constant pressure $p$ = 101325 Pa)
and the upper part is defined as a gas inlet with constant gas velocity of 0.2\,m/s (corresponding
to a constant flow of 56\,sccm). He (c=1) or N$_2$ (c=0) are introduced into the effluent region
through this boundary and allow to simulate different ambient atmospheres. The inflow of ambient
gas represents the "infinitely" larger reservoir of ambient air in our laboratory compared to the
small volume of the plasma jet. Additionally, the flow field in the effluent region is stabilized
by this inflow without formation of any vortexes in the numerical solution, which is a necessary
condition to obtain a converging solution. This additional gas inflow into the effluent region
could potentially influence the model results, but we have checked that selecting other gas
velocities (4 time lower or higher) do not affect the model results presented later.
\\The gas flow streamlines and a color map of the He concentration $c$ for the case of air (N$_2$) as
ambient gas are also shown in figure \ref{fig:geometry}. It can bee seen that air diffuses into the
helium effluent as the distance from the nozzle of the $\mu$-APPJ is increased. Additionally, the
results show that back diffusion of air into the plasma channel can be neglected in this modeled ideal case.

\subsection{Model of chemical kinetics}
When the gas flow and He concentration simulation is finished, the reactions and transport of ROS
can be simulated.
\\Again, the transport of given species \emph{i} (\emph{i}\,=\,O, O$_3$, O$_2(^1\Delta_g)$, or vibrationally excited
oxygen O$_2$(\emph{v})) is simulated as a diffusion-convection transport:
\begin{equation}
\frac{\partial n_i}{\partial t} + \nabla \cdot(-D_i\nabla n_i) = R_i - \textbf{u} \nabla n_i
\label{Model_D}
\end{equation}
with $D_i$ and $R_i$ being diffusion coefficient and production/loss term of a given particle due
to gas phase reactions, respectively. Diffusion coefficients of O and O$_3$ in He are
$D_{O\,in\,He}$ = 1.29$\times$10$^{-4}$\,m$^2$s$^{-1}$ and D$_{O_3\,in\,He}$ =
0.713$\times$10$^{-4}$\,m$^2$s$^{-1}$ \cite{Waskoenig2010a}. $D_{O_3\,in\,N_2}$=
0.137$\times$10$^{-4}$\,m$^2$s$^{-1}$ is the diffusion coefficient for O$_3$ in N$_2$ and has been
averaged over the two stated values by Massman \cite{Massman}. Since no diffusion coefficient for O
in N$_2$ could be found in literature, the diffusion coefficient for Ne in N$_2$ has been
calculated using the results of Chapman and Enskog \cite{Poling}. The masses of Ne and O are
similar and both are atoms. This supports the assumption that both atoms exhibit a similar
diffusion coefficient. A diffusion coefficient for Ne in N$_2$ of
D$_{Ne\,in\,N_2}$=0.338$\times$10$^{-4}$\,m$^2$s$^{-1}$ has been derived and used for the diffusion
of O in N$_2$. A gas temperature of 300\,K is assumed in the model. The values of the diffusion coefficients
of O$_2$ in He ($D_{O_2\,in\,He}$ = 5.51$\times$10$^{-5}$\,m$^2$s$^{-1}$) and in N$_2$
($D_{O_2\,in\,N_2}$ = 2.32$\times$10$^{-5}$\,m$^2$s$^{-1}$) are used for O$_2(^1\Delta_g)$ and
vibrationally excited oxygen O$_2$(\emph{v}). The dependence of $D_i$ on the He concentration c is
calculated in the same way as described in the previous section (formula \ref{Model_Diff}). The dissociation
degree of O$_2$ molecules in the plasma is less than 5\% under our experimental conditions.
Therefore, the O$_2$ density is not calculated in the model, but it is assumed to be constant
(concentration of 0.6\%) in the case of He as ambient gas and is calculated as n$_{O_2} =
[0.006+(1-c)\times0.2]\times n_0$ with $n_0 = 2.45\,\times\,10^{25}$\,m$^{-3}$ in the case of air as ambient gas.
\\The boundary conditions are selected as follows. The boundary condition at any wall is different
for stable O$_3$ (insulation/symmetry) and reactive or excited species (O, O$_2(^1\Delta_g)$,
O$_2$(\emph{v})). Oxygen atoms and O$_2$(\emph{v}) can recombine or be deexcited at the surface, which has to be
considered in the model. The Neumann's boundary condition considering the surface loss is therefore
used \cite{Schneider2011, Chantry1987}. A surface loss probability of $\beta$ = 10$^{-3}$ has been
chosen for atomic oxygen and for O$_2$(\emph{v}). Due to the long lifetime of O$_2$($^1\Delta_g$) \cite{Newman2000}, it has been assumed to be unreactive at the walls. Particles are introduced into the model by setting their density at the gas inlet into the plasma channel. They leave the
volume through the convection losses at the gas outlet in the effluent region.
\\Table \ref{tab:reactions1} contains the gas phase reactions, which are implemented in the model. These
reactions have been proposed by Jeong \emph{et al.}\cite{Jeong2000} and we use them in the
simulation called Model 1. No reactions with ions or electrons are assumed because their densities in the effluent are negligible \cite{Waskoenig2010}. Additionally, the reactions shown in table \ref{tab:reactions2} have been added to explain our
measurements with air as ambient atmosphere. The reactions R$_9$ - R$_{10}$ of Model 2 are motivated and explained later in the chapter Results. The model with the complete set of reactions from table \ref{tab:reactions1} and \ref{tab:reactions2} is called Model 2.
\begin{table}
\centering
\begin{tabular}[c]{c c c c}
\hline
& reactions & rate constants & ref \\
\hline
R$_1$ & O + O$_2$ + He $\rightarrow$ O$_3$ + He & $3.4\times10^{-46} (300/T_g)^{1.2}$\,m$^6$s$^{-1}$ & \cite{Stafford2004}\\
R$_2$ & O + O$_2$ + O$_2$ $\rightarrow$ O$_3$ + O$_2$ & $6\times10^{-46} (300/T_g)^{2.8}$\,m$^6$s$^{-1}$ & \cite{Stafford2004} \\
R$_3$ & O + O$_2$ + N$_2$ $\rightarrow$ O$_3$ + N$_2$ & $6\times10^{-46} (300/T_g)^{2.8}$\,m$^6$s$^{-1}$   & est. $^{\circledast}$ \\
R$_4$ & O + O + He $\rightarrow$ O$_2$ + He & $1\times10^{-45}$\,m$^6$s$^{-1}$ & \cite{Stafford2004}\\
R$_5$ & O + O + O$_2$ $\rightarrow$ O$_2$ + O$_2$ & $2.56\times10^{-46} (300/T_g)^{0.63}$\,m$^6$s$^{-1}$ & \cite{Stafford2004}\\
R$_6$ & O + O + N$_2$ $\rightarrow$ O$_2$ + N$_2$ & $2.56\times10^{-46} (300/T_g)^{0.63}$\,m$^6$s$^{-1}$  & est. $^{\circledast}$\\
R$_7$ & O + O$_3$ $\rightarrow$ O$_2$ + O$_2$ & $1.5\times10^{-17}$ exp$(-2250/T_g)$\,m$^3$s$^{-1}$ & \cite{Jeong2000a}\\
R$_8$ & O$_2$($^1\Delta_g$) + O$_3$ $\rightarrow$ O + 2\,O$_2$ & $6.01\times10^{-17}$ exp$(-2853/T_g)$\,m$^3$s$^{-1}$ & \cite{Jeong2000a}\\
\hline
\end{tabular}
\caption{Reactions of Model 1 with rate constants. ($^{\circledast}$ No literature values for three body reactions with N$_2$ have been found. Instead, the reaction rate of the corresponding reaction with O$_2$ as collision partner has been chosen, cf. \cite{Stafford2004}.)}
\label{tab:reactions1}
\end{table}
\begin{table}
\centering
\begin{tabular}[c]{c c c }
\hline
& reactions & rate or time constants  \\
\hline
R$_{9}$ & O$_2$(\emph{v}) + O$_2$ $\rightarrow$ O$_3$ + O & 2 $\times$ $10^{-21}$\,m$^3$s$^{-1}$ \\
R$_{10}$ & O$_2$(\emph{v}) $\rightarrow$ O$_2$ & 5 $\times$ $10^{-4}$\,s \\
\hline
\end{tabular}
\caption{Additional reactions of Model 2. Rate and time constants have been used as fitting parameters.}
\label{tab:reactions2}
\end{table}
\\The volume is again divided into two parts: the plasma channel, where steady state situation with
constant densities of all species is assumed, and effluent, where the reactive species can recombine.
This assumption is supported by two-photon absorption laser-induced fluorescence measurements
\cite{Knake2008}, which show a constant O atom density in the plasma channel and a decaying density in
the effluent. Therefore, no reactions are assumed in the plasma channel (the ROS densities are
introduced into the model via boundary condition). The reactions from table \ref{tab:reactions1} and \ref{tab:reactions2} are considered only
in they effluent region.
\\Table \ref{tab:parameters} shows the used boundary conditions and parameters of both models. The shown initial densities are introduced at the gas inlet of the plasma channel.
\begin{table}
\centering
\begin{tabular}[c]{c c c }
\hline
& value & note  \\
\hline
initial He density & $2.48\times10^{25}$ m$^{-3}$  & He at 101325\,Pa\\
initial O$_2$ density & $1.49\times10^{23}$\,m$^{-3}$  & 0.6\,\% of He density\\
O$_2$ content in He/air mixture & 0.2 $\cdot$ (1-c) & 20\,\% O$_2$ in air\\
initial O density & $8.5\times10^{20}$\,m$^{-3}$ & extrapolated value from \cite{Ellerweg2010}\\
initial O$_3$ density & $3.5\times10^{20}\,$m$^{-3}$ & extrapolated value from \cite{Ellerweg2010}\\
initial O$_2$(\emph{v}) density & $3\times10^{21}\,$m$^{-3}$ & fitting parameter\\
initial O$_2$($^1\Delta_g$) density & $7\times10^{20}\,$m$^{-3}$ & evaluated using \cite{Ellerweg2010} and \cite{Jeong2000a}\\
gas temperature T$_g$ & 300\,K & \\
\hline
\end{tabular}
\caption{Used parameters in the model}
\label{tab:parameters}
\end{table}

\section{Results}

The experimentally determined and simulated O and O$_3$ densities will be presented, compared and
discussed here. First, the measurements with He as ambient gas will be shown and
discussed. Afterwards, the results with ambient air will be presented. The results of Model 1 are discussed together with these measurements. Finally, the experiments performed to reveal the O$_3$ generation in the effluent will be discussed. Here, the reactions of Model 2 are also motivated and explained and the corresponding calculated densities are discussed.

\subsection{Validation of the model with He as ambient atmosphere}

\begin{figure}
    \centering
        \includegraphics[width=0.7\textwidth]{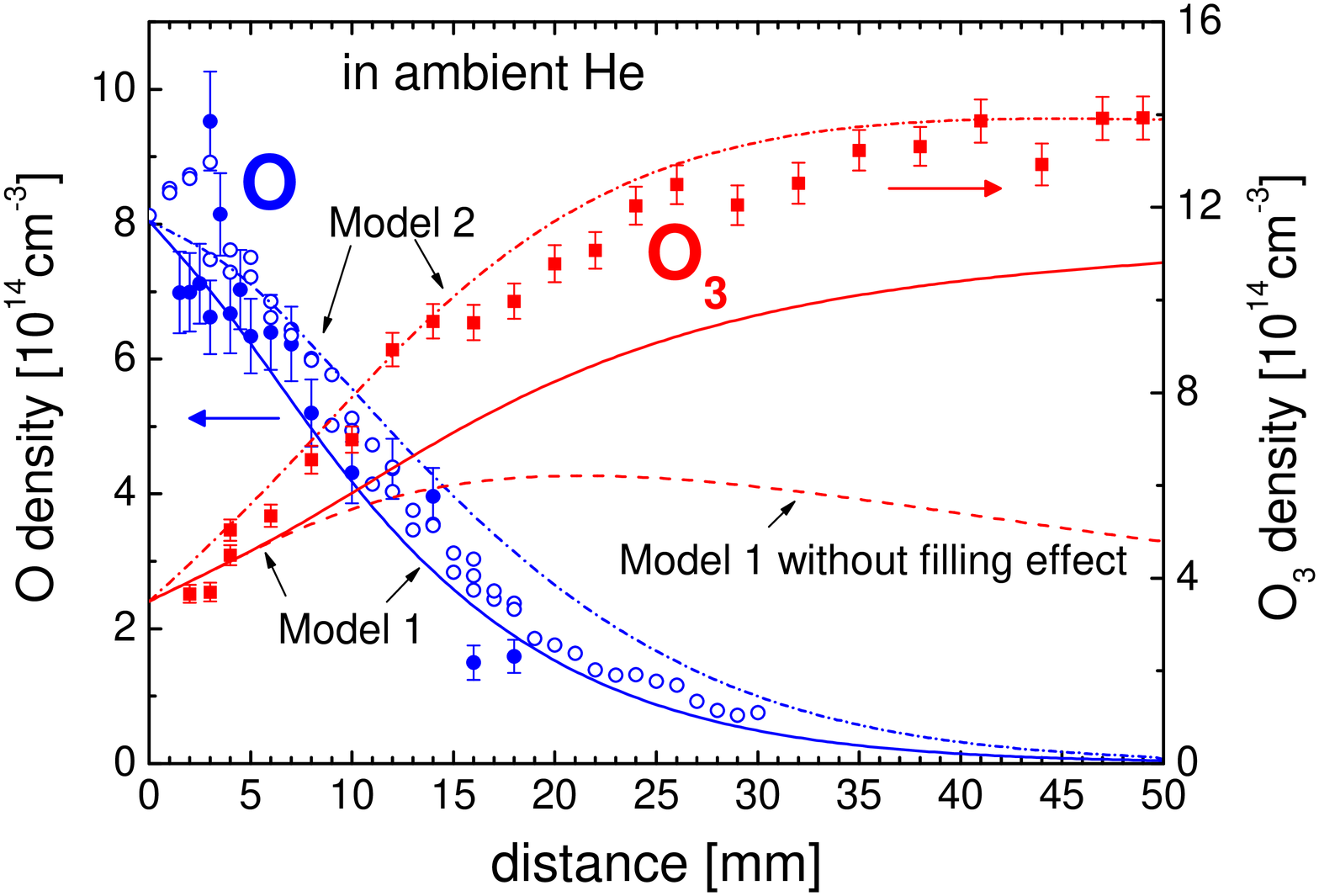}
    \caption{Atomic oxygen (measured by MBMS: \color{blue}$\medbullet$\color{black}, measured by TALIF: \color{blue}$\medcirc$\color{black}) and ozone (\color{red}$\blacksquare$\color{black}) density as a function of the distance from the jet in ambient helium (applied electrode voltage: 230\,V$_{RMS}$, gas flow: 1.4\,slm He with 0.6\,\% O$_2$). The lines reflect the densities derived by the Model 1 and 2. The TALIF measurement has been scaled down by a factor of 0.27. The results from MBMS and TALIF measurements are already published \cite{Ellerweg2010}.}
    \label{fig:he}
\end{figure}
Figure \ref{fig:he} shows the atomic oxygen and ozone density measured by MBMS and two-photon
absorption laser induced fluorescence spectroscopy (TALIF) in a helium ambiance as function of the distance. These
measurements have already been discussed within a previous publication \cite{Ellerweg2010}. In
summary, the maximum O density is in the range of 7$\times10^{14}\,$cm$^{-3}$ -
9$\times10^{14}\,$cm$^{-3}$ and declines from its maximum with the distance. The O$_3$ density is
3.7$\times10^{14}\,$cm$^{-3}$ at the nozzle of the $\mu$-APPJ and increases continuously with
increasing distance. A saturation value of 1.4$\times10^{15}\,$cm$^{-3}$ is reached beyond a
distance of 40\,mm. The atomic oxygen results obtained by MBMS and TALIF are in good agreement with
each other despite of a discrepancy by a factor of 3.7. The TALIF measurements reveal a O densities
3.7 times higher than the one obtained by MBMS. The origin of the discrepancy are possibly
uncertainties of the respective calibration processes. However, considering the fact that both measurements are performed by two different, independent diagnostics, this result can still be regarded as a very good agreement.
\\The atomic oxygen density simulated by the fluid model (Model 1) matches perfectly the MBMS and
TALIF measurements. In contrast, the simulated ozone density profile ("Model 1 without filling
effect" in figure \ref{fig:he}) doesn't correspond to the measurements. The modeled density
exhibits a maximum in a distance of 21\,mm and decreases beyond. For this model, helium is injected
through the gas inlet at the side. This condition reflects a situation with the plasma effluent
located in an infinitely large volume filled with He. However, the $\mu$-APPJ is located in a
chamber for the controlled He atmosphere with a volume of 1.1\,dm$^3$. This chamber fills
continuously with ozone during the time the plasma is operated and the ozone diffuses back into the
plasma effluent and causes an artificially higher ozone signal. To take this filling of the chamber
with ozone into account, we assume that all O recombines into O$_3$ (neglecting O losses due to
reactions R$_4$-R$_8$) and that an O$_3$ density of n$_{O,init}$ + n$_{O_3,init}$ is recycled
through the gas inlet in the effluent region (boundary condition at the gas inlet). This assumption
is reasonable, because we have tested the effect of the reactions R$_4$-R$_8$ on the model results
under our experimental conditions. The change of the densities is less than 5$\%$ when reaction R$_4$-R$_8$ are implemented. The gas phase chemistry is dominated by the O recombination with O$_2$ forming O$_3$ (R$_1$-R$_3$). The result of the simulation with this
recirculation of ozone is represented by the solid line in figure \ref{fig:he}. Now, the modeled
data reproduce the trend of the measurement very well, even if the absolute values do not match
perfectly. The profile of the O density is the same for both cases.
\\In summary, we state that Model 1 and the measurements are in reasonable agreement and that the
proposed chemistry of Jeong \emph{et al.} can indeed reasonably well describe the O and O$_3$
densities in the effluent in a helium ambiance. Here, the reactions R$_1$ - R$_3$ would even be sufficient to simulate
the effluent in an ambient helium atmosphere.

\subsection{Validation of the model with air as ambient atmosphere}

\begin{figure}
    \centering
        \includegraphics[width=0.6\textwidth]{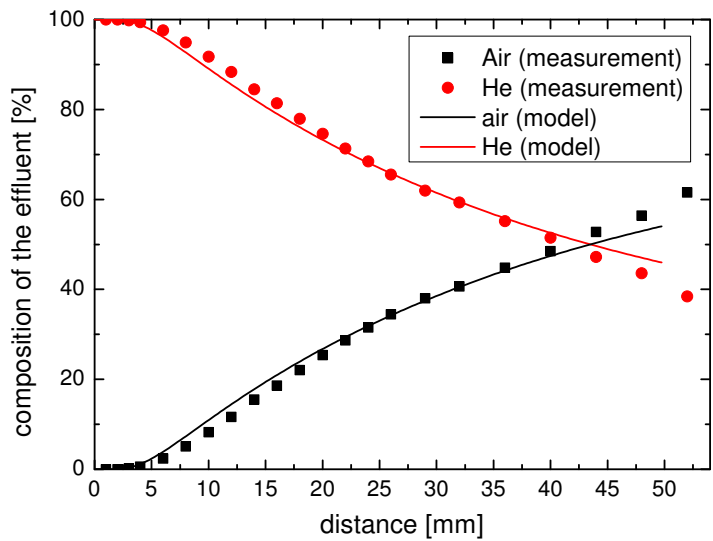}
    \caption{Composition of the effluent as a function of the distance (1.4\,slm helium flow)}
    \label{fig:air-admixture}
\end{figure}

A measurement of N$_2$ and O$_2$ signals as function of the distance has been performed in ambient air to determine
how fast air mixes into the helium effluent. The $\mu$-APPJ was just operated at a gas flow of
1.4\,slm helium without addition of oxygen and without igniting a plasma. The distance between the
$\mu$-APPJ and the sampling orifice was varied between 1 and 52\,mm while recording the He-, N$_2$-
and O$_2$ signal (cf. figure \ref{fig:air-admixture}). The sum of the N$_2$ and O$_2$ signal
determines the air signal. The shown results have been corrected for composition distortion by
measurements of different known He/air mixtures.
\\Almost no air admixes into the helium effluent up to a distance of 4\,mm. Air appears in the helium effluent beyond this point and its concentration
is increasing continuously as the distance is increased. The  gas composition on the symmetry axis,
as calculated in the fluid model, is also shown in figure \ref{fig:air-admixture}. The results from
the model are in a very good agreement with the measured data and corroborates that the fluid
model can very well simulate the admixture of air into the plasma effluent.

\subsection{Effluent chemistry in ambient air}

\begin{figure}
    \centering
        \includegraphics[width=0.6\textwidth]{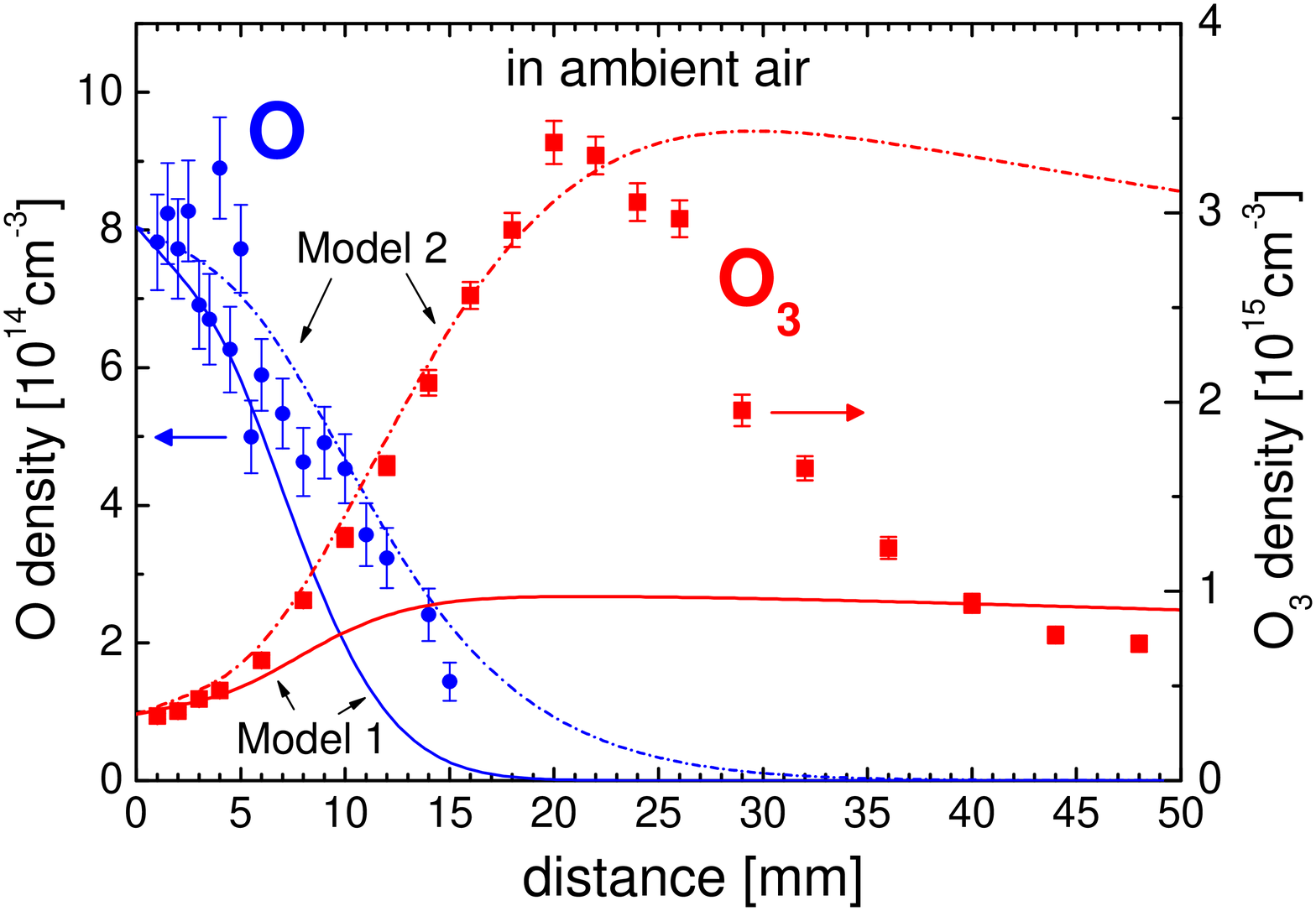}
    \caption{Atomic oxygen (\color{blue}$\medbullet$\color{black}) and ozone
    (\color{red}$\blacksquare$\color{black}) density as a function of the distance from the jet in
    ambient air (applied electrode voltage: 230\,V$_{RMS}$, gas flow: 1.4\,slm He with 0.6\,\%
    O$_2$). The solid lines reflect the densities derived by the model.} \label{fig:air}
\end{figure}

The MBMS measurements of the plasma effluent are now repeated with air as ambient gas. Again, the
densities of atomic oxygen and ozone are determined and are shown in figure \ref{fig:air}. The atomic
oxygen density is about 8$\times$10$^{14}$\,cm$^{-3}$ at the nozzle of the $\mu$-APPJ and declines
with increasing distance almost linearly. But still in a distance of 15\,mm oxygen atoms are
present in the effluent (ca. 1.5$\times$10$^{14}$\,cm$^{-3}$). The atomic oxygen trends obtained in
ambient air and helium are very similar. No TALIF measurements of O density are available for
comparison because the TALIF signal is influenced by additional non radiative
quenching of the excited states by air.
\\There is no difference between the ozone densities as measured in He and air atmosphere up to a
distance of 4\,mm. This is an expected result because the air starts to admix into the effluent
only beyond this distance (cf. figure \ref{fig:air-admixture}). At larger distances, the ozone
density in ambient air significantly increases in contrast to the density in ambient helium. A
maximum O$_3$ density of 3.4$\times$10$^{15}$cm$^{-3}$ is reached in a distance of 20\,mm.
Afterwards, the O$_3$ density decreases to a value of 0.7$\times$10$^{15}$cm$^{-3}$ at a distance
of 48\,mm. The steep decline of the O$_3$ density in ambient air beyond a distance of 20\,mm can be
explained by the effect of buoyancy force. Absorption spectroscopy measurements of ozone by H.
Bahre \emph{et al.} \cite{Bahre2010} have shown that the helium effluent (including the small
concentration of ozone molecules) starts to ascend in a surrounding air atmosphere beyond a
distance of ca. 20\,mm. The origin of buoyancy can be the lighter density of helium in contrast to
air combined with the slightly higher temperature of the plasma effluent. Later measurements of
ozone with the modified MBMS setup does not show such a distinct shape due to a different configuration. The jet is mounted vertically and the front plate with the sampling orifice is oriented horizontally at the modified MBMS setup.
\\We have seen in the previous case, that the conversion of all O into O$_3$ leads to a ozone density
of about 1.2 to 1.4$\times$10$^{15}$\,cm$^{-3}$. This means that more ozone is produced due to
admixture of air into the effluent than due to atomic oxygen emanating from the plasma. However,
there is no reaction in the proposed scheme of Jeong \emph{et al.}, which could explain the
additional production of ozone. This can be clearly seen, when the measured densities are compared
to the results of Model 1, also plotted in figure \ref{fig:air}.
\\The Model 1 predicts a faster decrease of the atomic oxygen density due to a faster reaction rate of reactions R$_1$, R$_2$, R$_3$, R$_5$, and R$_6$. This is caused by more available O$_2$ and N$_2$ in the effluent because of the air diffusion into it. Correspondingly, ozone is
produced faster compare to the case in ambient He. However, its maximal density predicted by Model
1 is still limited by the limited amount of oxygen atoms from the plasma and cannot exceed the
limit given by the sum of initial O and O$_3$ densities. Both O and O$_3$ densities simulated by
Model 1 do not reproduced the measured data (cf. figure \ref{fig:air}). The atomic oxygen density does not
decrease as fast as expected by the model, but stays relatively high even at a larger distance
from the jet and, as already mentioned, much more ozone is produced. These results indicate that
some reactions, which generate additional O atoms and O$_3$ molecules, are missing in the simulation Model 1.
\\The additional production of O atoms has already been observed by other authors. Reuter \emph{et
al.} \cite{Reuter2008} have observed that O atoms could be detected by TALIF even at the distance
of 10\,cm from a similar, but larger, planar jet operated in He with 0.5\,\% O$_2$ gas mixture and in He
as ambient atmosphere. They concluded that electrons and ions as well as metastable helium atoms
can be excluded as an origin for an energy transport into the effluent because their respective
densities quickly decrease, as soon as electric excitation ceases. Additionally, they have detected
energetic vacuum ultraviolet (VUV) radiation by OES and have shown that (V)UV radiation originating
from the discharge region reaches far into the effluent. Therefore, they have concluded that "(V)UV
radiation produces atomic oxygen in the effluent by dissociation of ozone or molecular oxygen."
\cite{Reuter2008} This conclusion has been corroborated in their following work, where the plasma
emitted photons have been blocked, which resulted in a decrease of O densities in the effluent
\cite{Reuter_thesis}. (V)UV radiation as a source of reactive species in the effluent of the
$\mu$-APPJ has also been observed indirectly during the inactivation of bacteria by the effluent of
the X-Jet \cite{Schneider2011,Schneider2011b}. We will try now, with the help of the absolute
calibrated data, additional MBMS measurements, and adjustment of the fluid model, to
reveal more details about a possible O and O$_3$ generation scheme in the plasma effluent.

\subsection{Revealing the origin of the additional O and ozone}
\begin{figure}
    \centering
        \includegraphics[width=0.6\textwidth]{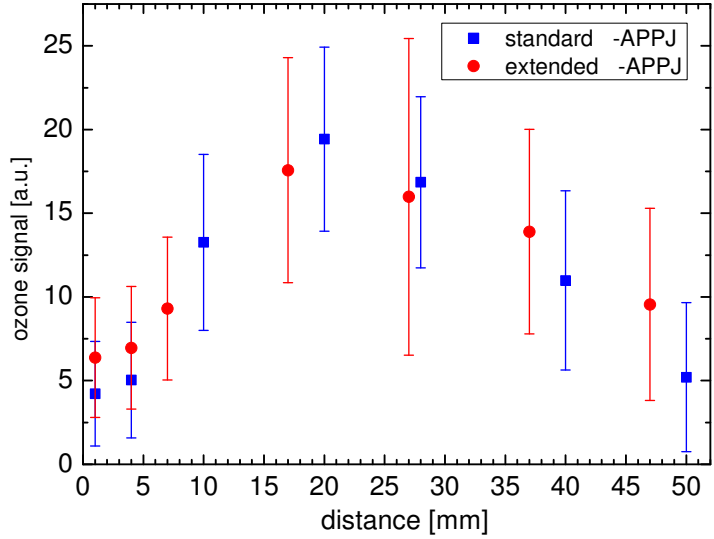}
    \caption{Relative MS signal at mass 48 amu (ozone) as function of the distance to the nozzle
    for the standard $\mu$-APPJ and extended $\mu$-APPJ under same conditions (in ambient air,
    applied electrode voltage: 230\,V$_{RMS}$, gas flow: 1.4\,slm He with 0.6\,\% O$_2$).}
    \label{fig:extendedappj}
\end{figure}

\emph{The possible effect of air back diffusion}\\
As already discussed, the results in the literature indicate that plasma generated (V)UV photons
can initiate photodissociation reactions leading to formation of additional O and subsequently
O$_3$. Before we, however, discuss the possible photodissociation reactions, we need to check for
the possible effect of air back diffusion into the plasma channel. The gas flow simulation has
shown, that air back diffusion is almost negligible. But the model represents an ideal case without
effects due to actual design and realization of the plasma jet such as diffusion along the corner
edge of the square geometry. Additionally, the simulation shows that even in the ideal case the air
can come into direct contact with the active plasma zone. Point A in figure \ref{fig:geometry}
marks the edge of the plasma channel, where it takes place. The electron driven dissociation of
atmospheric O$_2$ followed by the diffusion of the products towards the jet axis could also be a
source of additional O and O$_3$. To exclude this effect, we have built a modified $\mu$-APPJ with
an elongated gas channel. This gas channel is made from glass plates, starts directly at the
electrodes, has a length of 3\,mm and a 1$\times$1\,mm$^2$ cross section. The plasma effluent is
therefore confined the first 3\,mm after the plasma region and a direct interaction between
atmospheric air and the plasma is prevented. Figure \ref{fig:extendedappj} shows the comparison of
the ozone signals obtained with the standard $\mu$-APPJ without extension and the extended
$\mu$-APPJ. Only relative data are shown, no absolute calibration and compensation for the changing
composition distortion have been performed. No significant difference can be found between both
measurements. One can, therefore, conclude that diffusion of air into the plasma cannot explain the
increased ozone (and also O) production.
\\During these measurements, the newer MBMS setup with horizontally oriented front plate has been
used. Therefore, the influence of the helium buoyancy is different and the shape of the measured
ozone signal differs a little bit from the previous measurement (cf. figure \ref{fig:air}).\\

\emph{Photodissociation as the source of additional O and O$_3$}\\
Photodissociation of oxygen molecules or ozone has been suggested as a possible source of
additional O and consequently also O$_3$. The quantitative and space resolved MBMS measurements of O$_3$ allows now the quantitative estimation of the necessary O$_3$ production rate needed to
explain the rise of its density as observed in figure \ref{fig:air}. This estimated production rate
can be used further to estimated the necessary photon flux, if the photodissociation is the main
source of this ozone. The O$_3$ density rises from 0.5$\times10^{15}$ to 3.0$\times10^{15}\,\mbox{cm}^{-3}$
on the distance of 13\,mm (starting at 6 mm from the jet nozzle). The fluid model predicts the
average gas velocity on the axis of about 25\,m/s resulting in the transport time t=0.013/25\,s =
0.52\,ms. The averaged ozone production rate in the effluent can therefore be estimated as:
\begin{equation}
\frac{\partial n_{O_3}}{\partial t}=2.5\times10^{15}\,\mbox{cm}^{-3}/5.2\times10^{-4}\,\mbox{s}=4.8\times10^{18}\,\mbox{cm}^{-3}\mbox{s}^{-1}
\sim 2\cdot\Gamma \cdot \sigma_{ph} \cdot n_{source}.\label{ozone_rate}
\end{equation}
The right hand side of equation \ref{ozone_rate} is the estimated production rate of the ozone due
to photodissociation of some source molecule (e.g. O$_2$ or O$_3$) with density $n_{source}$. The
$\sigma_{ph}$ is the photodissociation cross section, $\Gamma$ is the photon flux from plasma per
unit area and second, and factor 2 takes into account that two O$_3$ molecules can be produced  in
each case. The possible reactions could be the direct dissociation of O$_2$ into 2\,O atoms, which
than quickly react with O$_2$ from the air into O$_3$ or the photodissociation of O$_3$ into O and
highly vibrationally excited O$_2$(\emph{v}), which can again react with O$_2$ producing O$_3$ and O in
this case. The two O atoms will as in the previous case react to O$_3$ providing again a net
production of two new O$_3$ molecules.
\\The typical maximum of the photodissociation cross sections of O$_2$ or O$_3$ molecules ($\sigma_{ph}\sim 10^{-17}\,\mbox{cm}^{2}$) \cite{Voigt2001, Baloitcha2006}, can be used to estimate the lowest necessary photon flux in the ideal case, in which each photon is absorbed with maximum probability. With n$_{O_2}\sim10^{24}\,\mbox{m}^{-3}$ and n$_{O_3}\sim2\times10^{21}\,\mbox{m}^{-3}$ the photon fluxes of at least
$\Gamma_{O_2}\sim2.5\times10^{21}\,\mbox{m}^{-2}\mbox{s}^{-1}$ and $\Gamma_{O_3}\sim1.2\times10^{24}\,\mbox{m}^{-2}\mbox{s}^{-1}$ are
necessary for the ozone production via photodissociation of O$_2$ or O$_3$, respectively. With the 1
mm$^2$ cross-section area of the jet and a chosen photon energy of 5\,eV, we obtain a necessary
energy flux of at least 2\,mW in the case of dissociation of O$_2$ and 1\,W in the case of
photodissociation of O$_3$. The latter reaction can be directly excluded, because the power
absorbed by the plasma is for sure lower than 1\,W and the plasma cannot generate so many photons.
Even in the case of photodissociation of O$_2$ seems the power of the photon flux too high,
considering the crude simplifications in the estimation of the photon flux. However, it could still
contribute to the production of O and ozone in the effluent.\newline
\\Luckily, the photodissociation reaction of O$_2$ by plasma generated photons can directly be tested
by the X-Jet modification of the $\mu$-APPJ. For this experiment the additional He flow with
different concentrations of O$_2$ is flown through the side channel of the X-Jet, whereas the
plasma is ignited in the standard mixture of He with 0.6\% of O$_2$ in the direct channel between
electrodes. The additional flow diverts the plasma effluent from the direct channel into the side
channel after the crossing of both channels and fills at the same time the part of the direct
channel after the crossing with a He/O$_2$ gas mixture as shown in the figure \ref{fig:x-jet}a.
The plasma generated photons, which are propagating through the direct channel of the X-Jet,
overlap spatially with the He/O$_2$ gas. The possible flow of O and O$_3$ from the direct channel,
generated now by photodissociation reactions, can be measured by the MBMS system. O$_2$
concentrations of up to 20\% and different distances between the nozzle of the direct channel and
the sampling orifice of the MBMS were tested to allow the photons traveling different lengths
through the He/O$_2$ mixture and also to allow air to diffuse into the gas flow. Ozone has never
been detected in these measurements. Additionally, the normal $\mu$-APPJ operated in pure He gas,
without addition of O$_2$ gas, has been operated in ambient atmosphere and the production of ozone
in the effluent at different distances from the jet has been measured. Even more VUV radiation is
expected to be produced in this case due to slower quenching of excited states and higher densities
of helium excimers. The ozone density was also in this case below the detection limit of the MS.
All these measurements are evidence that photodissociation of ground state O$_2$ molecules can be
very probably excluded as the source of additional O and O$_3$ in the effluent. It seems, that
photodissociation reactions are not the source of additional O atoms and ozone molecules under our
experimental conditions and some other oxygen containing species generated in the plasma have to be
involved in this process.\newline
\\Summarizing the facts discussed up to now, the following conditions have to be fulfilled in the
involved reaction mechanism: (i) the VUV radiation can be probably excluded as a source of
additional O and O$_3$ due to the above discussed reasons; (ii) molecular ground state oxygen from
air is required because the additional ozone signal only appears when air diffuses into the
effluent; (iii) atomic oxygen is produced in this reaction because the O densities are at larger
distances from the jet higher as predicted by Model 1 (cf. figure\ref{fig:air}); (iv) some reactive
oxygen species from the plasma have to be involved in the reaction because the interaction of the
effluent of pure He plasma with ambient air does not lead to formation of O and O$_3$; and (v) the
reaction has to be somehow limited by the O$_2$ density because the effect is weak with 0.6\% of
O$_2$ in the mixture (the measurements with He as ambient atmosphere) but becomes significant at
higher O$_2$ densities when air admixes into the effluent.\newline
\\The reaction of the highly vibrationally excited O$_2$(\emph{v}) molecules, which are generated in the
plasma and have a limited life time, with the ground state O$_2$ is proposed here as a possible
reaction mechanism. The vibrationally excited oxygen molecules can directly react with ground
state oxygen molecules (reaction R$_{9}$) producing an oxygen atom and an ozone molecule. Two O$_3$
are therefore effectively formed per reaction. The reactions of highly excited O$_2$(\emph{v}$>$26)
molecules have been for example discussed in the literature and were proposed to solve the ozone
deficit problem in the atmosphere \cite{Miller1994,Flynn1996}. Additionally, the O$_2$(\emph{v}) relax
decay to its ground state (reaction R$_{10}$) is proposed to explain the small effect observed in
the experiments in ambient He. This reaction scheme has been included to the simulation (Model
2). Since some of the needed parameters are unknown, they have been adjusted until Model 2 fits
reasonably well the measured densities. The fitting parameters are reaction rate of reaction
R$_{9}$ and the lifetime and the initial density of the vibrationally excited oxygen molecules. The
used values of the fitting parameters are shown in the tables \ref{tab:reactions2} and
\ref{tab:parameters}. The results of the Model 2 are shown in figure \ref{fig:he} for the case of
ambient helium and in figure \ref{fig:air} for the case of ambient air. Model 2 is in very good
agreement with the measurements for both cases, ambient helium and ambient air.
\\This result corroborates that a reasonable reaction mechanism without involvement of photons can
explain under our experimental conditions the observed results. We stress here, that it is just a
hypothesis. A validation by possible measurements of O$_2$(\emph{v}) and by comparison of fitted rates and
relaxation times with measured rates and relaxation times, if they exist, should be performed
before accepting this reaction mechanism. Additionally, it should be checked whether this proposed
reaction mechanism can explain the observations, which have lead to the conclusion that VUV and UV
photons are involved in the effluent chemistry.

\section{Conclusion}

Molecular beam mass spectrometry is used for measurements of absolute densities of O and O$_3$
in the effluent of a He/O$_2$ micro-scaled atmospheric pressure plasma jet operated in ambient air
and ambient helium. The effect of the composition distortion in the molecular beam is studied and
carefully considered in the calibration process. Additionally, the admixture of air into the
plasma effluent has been quantified. A simple fluid model of the gas flow, admixture of the
air into the effluent, and reaction kinetics of several reactive oxygen species is constructed and
used to test different reaction mechanisms.
\\It was observed that the admixture of air into the effluent leads to an additional production of
oxygen atoms and a fast increase of ozone density. Surprisingly, two to three times more ozone is
produced in the additional reactions of the plasma effluent with the ambient air than what is
observed in ambient helium without big differences in the distance variation of atomic oxygen
density. This observation can have big implication for the use of similar jets in the plasma
medicine and plasma sterilization applications. The possible role of plasma generated photons has
been carefully tested. It is shown that the photodissociation of O$_2$ and O$_3$ is not responsible
for the observed effect and that the photon flux is very probably too low to have any effect. A
reaction scheme involving the reaction of plasma produced highly vibrationally excited oxygen
molecules with a limited lifetime with ground state O$_2$ molecules is proposed as a possible
explanation of the observed O and O$_3$ densities. A very good agreement between measured and
simulated densities is achieved.

\ack
This project is supported by DFG within the framework of the Research Group FOR1123 (project C1) and with the individual grant KE 1145/1-1 and by the Research Department 'Plasmas with Complex Interactions'.

\section*{References}
\bibliographystyle{unsrt}

\begin{thebibliography}{10}

\bibitem{Schafer2008}
J.~Sch\"{a}fer, R.~Foest, A.~Quade, A.~Ohl, and K.-D. Weltmann.
\newblock {Local deposition of SiOx plasma polymer films by a miniaturized
  atmospheric pressure plasma jet (APPJ)}.
\newblock {\em Journal of Physics D: Applied Physics}, 41(19):194010, 2008.

\bibitem{Benedikt2007}
J.~Benedikt, V.~Raballand, A.~Yanguas-Gil, K.~Focke, and A.~von Keudell.
\newblock {Thin film deposition by means of atmospheric pressure microplasma
  jet}.
\newblock {\em Plasma Physics and Controlled Fusion}, 49(12B):B419--B427, 2007.

\bibitem{Raballand2009}
Vanessa Raballand, Jan Benedikt, Sven Hofmann, Max Zimmermann, and Achim von
  Keudell.
\newblock {Deposition of silicon dioxide films using an atmospheric pressure
  microplasma jet}.
\newblock {\em Journal of Applied Physics}, 105(8):083304, 2009.

\bibitem{Goree2006}
J~Goree, B~Liu, and D~Drake.
\newblock {Gas flow dependence for plasma-needle disinfection of S-mutans
  bacteria}.
\newblock {\em J. Phys. D: Appl. Phys.}, 39:3479, 2006.

\bibitem{Rahul05}
R~Rahul, O~Stan, A~Rahman, E~Littlefield, K~Hoshimiya, A~P Yalin, A~Sharma,
  A~Pruden, C~A Moore, Z~Yu, and G~J Collins.
\newblock {Optical and RF electrical characteristics of atmospheric pressure
  open-air hollow slot microplasmas and application to bacterial inactivation}.
\newblock {\em J. Phys. D}, 38:1750, 2005.

\bibitem{Perni2007}
S~Perni, G~Shama, J~L Hobman, P~A Lund, C~J Kershaw, G~A Hidalgo-Arroyo, C~W
  Penn, X~T Deng, J~L Walsh, and M~G Kong.
\newblock {Probing bactericidal mechanisms induced by cold atmospheric plasmas
  with Escherichia coli mutants}.
\newblock {\em Applied Physics Letters}, 90:73902, 2007.

\bibitem{Sladek2005a}
R.~E.~J. Sladek and E.~Stoffels.
\newblock {Deactivation of Escherichia coli by the plasma needle}.
\newblock {\em Journal of Physics D: Applied Physics}, 38(11):1716--1721, 2005.

\bibitem{Chiang2007}
W.~Chiang and R.~M. Sankaran.
\newblock {Microplasma synthesis of metal nanoparticles for gas-phase studies
  of catalyzed carbon nanotube growth}.
\newblock {\em Applied Physics Letters}, 91(12):121503, 2007.

\bibitem{Walsh2010}
J.~L. Walsh, F.~Iza, N.~B. Janson, V.~J. Law, and M.~G. Kong.
\newblock {Three distinct modes in a cold atmospheric pressure plasma jet}.
\newblock {\em Journal of Physics D: Applied Physics}, 43(7):075201, 2010.

\bibitem{Knake08}
N~Knake, S~Reuter, K~Niemi, V~{Schulz-von der Gathen}, and J~Winter.
\newblock {Absolute atomic oxygen density distributions in the effluent of a
  microscale atmospheric pressure plasma jet}.
\newblock {\em J. Phys. D: Appl. Phys.}, 41:194006, 2008.

\bibitem{Karakas2010}
E.~Karakas, M.~Koklu, and M.~Laroussi.
\newblock {Correlation between helium mole fraction and plasma bullet
  propagation in low temperature plasma jets}.
\newblock {\em Journal of Physics D: Applied Physics}, 43(15):155202, 2010.

\bibitem{Reuter2011}
R.~Reuter, D.~Ellerweg, A.~von Keudell, and J.~Benedikt.
\newblock {Surface reactions as carbon removal mechanism in deposition of
  silicon dioxide films at atmospheric pressure}.
\newblock {\em Applied Physics Letters}, 98(11):111502, 2011.

\bibitem{Schneider2011}
S.~Schneider, J.-W. Lackmann, F.~Narberhaus, J.~E. Bandow, B.~Denis, and
  J.~Benedikt.
\newblock {Separation of VUV/UV photons and reactive particles in the effluent
  of a He/O 2 atmospheric pressure plasma jet}.
\newblock {\em Journal of Physics D: Applied Physics}, 44(37):379501, 2011.

\bibitem{Knake2008}
N~Knake, K~Niemi, S~Reuter, V~{Schulz-von der Gathen}, and J~Winter.
\newblock {Absolute atomic oxygen density profiles in the discharge core of a
  microscale atmospheric pressure plasma jet}.
\newblock {\em Applied Physics Letters}, 93:131503, 2008.

\bibitem{Ellerweg2010}
Dirk Ellerweg, Jan Benedikt, Achim von Keudell, Nikolas Knake, and Volker
  {Schulz-von der Gathen}.
\newblock {Characterization of the effluent of a He/O 2 microscale atmospheric
  pressure plasma jet by quantitative molecular beam mass spectrometry}.
\newblock {\em New Journal of Physics}, 12(1):013021, January 2010.

\bibitem{Sousa2008}
J.~S Sousa, G.~Bauville, B.~Lacour, V.~Puech, M.~Touzeau, and L.~C. Pitchford.
\newblock {O2 (a1$\Delta$g) production at atmospheric pressure by
  microdischarge}.
\newblock {\em Applied Physics Letters}, 93(1):011502, 2008.

\bibitem{Benedikt2009}
Jan Benedikt, Dirk Ellerweg, and Achim von Keudell.
\newblock {Molecular beam sampling system with very high beam-to-background
  ratio: the rotating skimmer concept.}
\newblock {\em The Review of scientific instruments}, 80(5):055107, May 2009.

\bibitem{Schneider2011b}
S.~Schneider, J.-W. Lackmann, D.~Ellerweg, B.~Denis, F.~Narberhaus, J.E.
  Bandow, and J.~Benedikt.
\newblock {The role of VUV radiation in the inactivation of bacteria with an
  atmospheric pressure plasma jet}.
\newblock {\em submitted to Plasma Processes and Polymers}, 2011.

\bibitem{Scoles1988a}
G.~Scoles.
\newblock {\em {Atomic and Molecular Beam Methods Vol. I}}.
\newblock Oxford Univ. Press, New York, 1988.

\bibitem{Knuth95}
E~L Knuth.
\newblock {Composition distortion in MBMS sampling}.
\newblock {\em Combustion and Flame}, 103:171, 1995.

\bibitem{Jeong2000}
J~Y Jeong, J~Park, I~Henins, S~E Babayan, V~J Tu, G~S Selwyn, G~Ding, and R~F
  Hicks.
\newblock {Reaction chemistry in the Afterglow of an oxygen-helium
  atmospheric-pressure plasma}.
\newblock {\em J. Phys. Chem. A}, 104:8027--8032, 2000.

\bibitem{Reuter2008}
S.~Reuter, K.~Niemi, V.~{Schulz-von der Gathen}, and H.~F. D\"{o}bele.
\newblock {Generation of atomic oxygen in the effluent of an atmospheric
  pressure plasma jet}.
\newblock {\em Plasma Sources Science and Technology}, 18(1):015006, 2008.

\bibitem{Brokaw1968}
R.~S. Brokaw.
\newblock {Viscosity of gas mixtures}.
\newblock Technical report, NASA, Washington D.C., 1968.

\bibitem{Pavlin}
T.~Pavlin.
\newblock {\em {Hyperpolarized Gas Polarimetry and Imaging at Low Magnetic
  Field}}.
\newblock PhD thesis, California Institute of Technology, 2003.

\bibitem{Poling}
B.~E. Poling and J.~M. Prausnitz.
\newblock {\em {The Properties of Gases and Liquids}}.
\newblock Mc-Graw Hill, 5 edition, 2001.

\bibitem{Waskoenig2010a}
J.~Waskoenig, K.~Niemi, N.~Knake, L.~M. Graham, S.~Reuter, V.~{Schulz-von der
  Gathen}, and T.~Gans.
\newblock {Atomic oxygen formation in a radio-frequency driven
  micro-atmospheric pressure plasma jet}.
\newblock {\em Plasma Sources Science and Technology}, 19(4):045018, 2010.

\bibitem{Massman}
W.~J. Massman.
\newblock {A Review of the Molecular Diffusivities of H2O , CO2 , CH4 , CO, O3
  , SO2 , NH3 , N2O , NO , AND NO2 in Air , O2 and N2 near STP}.
\newblock {\em Atmospheric Environment}, 32(6):1111--112, 1998.

\bibitem{Chantry1987}
P.~J. Chantry.
\newblock {A simple formula for diffusion calculations involving wall
  reflection and low density}.
\newblock {\em Journal of Applied Physics}, 62(4):1141, 1987.

\bibitem{Newman2000}
Stuart~M Newman, Andrew~J Orr-Ewing, David~A Newnham, and John Ballard.
\newblock {Temperature and Pressure Dependence of Line Widths and Integrated
  Absorption Intensities for the O 2 a 1 $\Delta$ g - X 3 $\Sigma$ g - (0,0)
  Transition}.
\newblock {\em The Journal of Physical Chemistry A}, 104(42):9467--9480,
  October 2000.

\bibitem{Waskoenig2010}
J~Waskoenig, K~Niemi, N~Knake, L~M Graham, S~Reuter, V~{Schulz-von der Gathen},
  and T~Gans.
\newblock {No Title}.
\newblock {\em Plasma Sources Sci. Technol.}, 19:45018, 2010.

\bibitem{Stafford2004}
D.~S. Stafford and M.~J. Kushner.
\newblock {O2 (1 delta) production in HeO2 mixtures in flowing low pressure
  plasmas}.
\newblock {\em Journal of Applied Physics}, 96(5):2451, 2004.

\bibitem{Jeong2000a}
James~Y. Jeong, Jaeyoung Park, Ivars Henins, Steve~E. Babayan, Vincent~J. Tu,
  Gary~S. Selwyn, Guowen Ding, and Robert~F. Hicks.
\newblock {Reaction Chemistry in the Afterglow of an Oxygen-Helium,
  Atmospheric-Pressure Plasma}.
\newblock {\em The Journal of Physical Chemistry A}, 104(34):8027--8032, August
  2000.

\bibitem{Bahre2010}
H.~Bahre.
\newblock {\em {Spatially resolved spectroscopic investigation of ozone
  formation in a miniaturised atmospheric pressure plasma jet}}.
\newblock Diploma thesis, Ruhr-Universit\"{a}t Bochum, 2010.

\bibitem{Reuter_thesis}
S.~Reuter.
\newblock {\em {Formation mechanism of atomic oxygen in an atmospheric pressure
  plasma jet characterised by spectroscopic methods}}.
\newblock Phd thesis, Universit\"{a}t Duisburg-Essen, 2008.

\bibitem{Voigt2001}
S.~Voigt, J.~Orphal, K.~Bogumil, and J.~P. Burrows.
\newblock {The temperature dependence (203-293 K) of the absorption cross
  sections of O3 in the 230-850 nm region measured by Fourier-transform
  spectroscopy}.
\newblock {\em Journal of Photochemistry and Photobiology A: Chemistry},
  143(1):1--9, 2001.

\bibitem{Baloitcha2006}
E.~Baloitcha and G.~Balint-Kurti.
\newblock {Theory of the photodissociation of ozone in the Hartley continuum;
  effect of vibrational excitation and O(1D) atom velocity distribution}.
\newblock {\em Phys. Chem. Chem. Phys.}, 7:3829--3833, 2005.

\bibitem{Miller1994}
R.L. Miller, a~G Suits, P~L Houston, R~Toumi, J~a Mack, and a~M Wodtke.
\newblock {The "Ozone Deficit" Problem: O2(X, v ge 26) + O(3P) from 226-nm
  Ozone Photodissociation.}
\newblock {\em Science (New York, N.Y.)}, 265(5180):1831--8, September 1994.

\bibitem{Flynn1996}
G.W. Flynn, C.S. Parmenter, and A.M. Wodtke.
\newblock {Vibrational energy transfer}.
\newblock {\em J. Phys. Chem.}, 100:12817--12838, 1996.

\end{thebibliography}

\end{document}